# Probabilistic Forecasting of Patient Waiting Times in an Emergency Department

Siddharth Arora[†], James W. Taylor[*] and Ho-Yin Mak[§]

Saïd Business School,
University of Oxford, Park End Street, Oxford, OX1 1HP, U.K.

[†] Siddharth Arora
Email: Siddharth.Arora@sbs.ox.ac.uk

[*]James W. Taylor
Email: James.Taylor@sbs.ox.ac.uk

[§]Ho-Yin Mak
Email: Ho-Yin.Mak@sbs.ox.ac.uk

# Probabilistic Forecasting of Patient Waiting Times in an Emergency Department

───────────────────────────────────────────────────────────────────

**Problem definition**: We study the estimation of the probability distribution of individual patient waiting times in an emergency department (ED). Our feature-rich modelling allows for dynamic updating and refinement of waiting time estimates as patient- and ED-specific information (e.g., patient condition, ED congestion levels) is revealed during the waiting process. Aspects relating to communicating forecast uncertainty to patients, and implementing this methodology in practice, are also discussed.

**Academic/Practical Relevance**: While it is known that waiting time estimates can help improve patients' overall satisfaction and prevent abandonment, existing methods focus on point forecasts. By providing personalized probabilistic forecasts, our approach gives patients and first responders a more comprehensive picture of the possible waiting trajectory, and provides more reliable inputs to inform prescriptive modelling of ED operations. For example, we demonstrate that publishing probabilistic waiting time estimates can inform patients in selecting an ED from a network of EDs, which can lead to more uniform spread of patient load across the network.

**Methodology**: We use the machine learning approach of quantile regression forest (QRF) to produce the probabilistic forecasts. Using a large patient-level dataset we extract the following categories of predictor variables: (1) calendar effects, (2) demographics, (3) staff levels, (4) ED workload due to patient volumes, and (5) the severity of patient condition. Rankings of predictor importance are derived from regression trees.

**Results**: The proposed approach generates more accurate probabilistic and point forecasts, when compared with methods proposed in the literature for modelling waiting times, and rolling average benchmarks typically used in practice. Patient workflow and calendar effects were identified as key predictors of ED waiting times.

**Managerial Implications**: For emergency healthcare service providers, probabilistic waiting time estimates could assist in ambulance diversion, staff allocation, and managing patient-flow, which could facilitate efficient operations and cost-savings while aiding better patient care and outcomes.

───────────────────────────────────────────────────────────────────

# 1. Introduction

Emergency departments (EDs) are coming under increasing pressure to provide safe and quality care to patients in a timely manner. It was estimated that between 2006 and 2016, there were 1.4 billion ED visits in the US alone, whereby the number of visits increased by around 2.3 million per year (Singer et al. 2019). The hospital staffing and infrastructure has not grown at the same rate, which has resulted in longer waiting times. In England, the national health service (NHS) has a pledge set out in the handbook of its constitution stating that 95% of patients attending the ED should be treated, admitted or discharged within four-hours (NHS 2019). However, for the first time in 2019, all major ED units in England missed their waiting time targets; while in the US, from 2003 to 2009, the mean waiting time increased by 25% (Hing and Bhuiya 2012). This is a matter of growing concern, as long ED waiting times are associated with increased morbidity and mortality, and are one of the leading causes of patient dissatisfaction (Bernstein et al. 2009). Waiting times and perceived queue length can influence patients to drop out from the ED (Batt and Terwiesch 2015). Patients that drop out from the ED are linked with having a higher likelihood of re-presentation and poor outcomes (Carter et al. 2014). Providing delay estimates can help lower the perceived waiting time (Jouini et al. 2011). Growing the capacity in EDs to eradicate congestion and minimize waiting times requires long-term planning and funding; meanwhile, providing patients with an estimate of their waiting times can be an inexpensive and immediate way forward to managing patient expectations and reducing abandonment rates.

For emergency healthcare service providers, long waiting times can have significant economic implications. In private healthcare systems (e.g., in the US), shorter and more transparent waiting times could lead to higher revenues for hospitals, since about 10% of total US healthcare cost is spent on emergency care (Galarraga and Pines 2016). In many OECD countries with public healthcare systems, service providers incur performance fines, including financial penalties, if they exceed the waiting time targets (OECD Executive Summary 2013). It is estimated that prolonging waiting time in the ED by just 10 minutes increases the cost of care by an average of 6% for a high-acuity patient, and 3% for a low-acuity patient (Woodworth and Holmes 2020). To deal with long waiting times (or congestion), EDs sometimes rely on external agencies to provide temporary workforce. However, temporary staff cost 20% more on average than the permanent staff (Buchan et al. 2019). Waiting time estimates could potentially assist hospitals in making more informed staffing decisions, thereby reducing the dependency on costly surge capacities. Given the impact of long waiting times on patient outcomes and its economic implications for service providers; streamlining patient-flow, minimizing waiting times, and optimizing resource allocation, while providing quality care, is at the heart of reforming emergency healthcare services.

Modelling of ED operations has been an active area of research in operations management. A significant stream of literature takes descriptive and prescriptive views on patient flows in the ED by use of queueing models (Armony 2015; Batt and Terwiesch 2015; Bayati 2017; Xu and Chan 2016), as well as discrete-event simulations (Baril et al. 2019). For a detailed review of literature in this area, see, for example (Hu et al. 2018; Keskinocak and Savva 2020; Misic and Perakis 2020; Singh and Terwiesch 2012), and references therein.

The literature on predicting patient waiting times in the ED is relatively scarce. Accurate predictive models for waiting times can provide valuable information to patients, assist service providers in planning and operations, as well as support prescriptive studies of ED operations (e.g., for calibrating queueing and simulation models). Ang et al. (2016) generate point forecasts for ED waiting times (from registration to start-of-treatment) based on the least absolute shrinkage operator (Lasso) using predictor variables inspired by fluid model estimators. They report an over 30% reduction in mean squared error, compared to a rolling average model, which is the standard method adopted in practice in US hospitals (Dong et al. 2019). Ding et al. (2010) generate estimates of treatment, boarding, and waiting room time using quantile regression, focussing on 10, 50, and 90% quantiles. Using data available at triage, Sun et al. (2012) use quantile regression to predict the median and 95% quantile of the waiting time (from triage to consultation). We are, however, not aware of any existing study that models and evaluates the whole probability distribution of ED waiting times.

Our proposed approach differs from the aforementioned studies in that it is: (1) probabilistic, and (2) personalized. Waiting time is inherently uncertain and the distribution is asymmetric (see Section 2.2), and so a point forecast can be uninformative and even potentially misleading, which could risk greater dissatisfaction among patients. Given that patients can feel increasingly dissatisfied if they end up waiting longer than the published estimate, it is imperative that the uncertainty associated with forecasts is adequately conveyed to the patients, so that patient expectations can be better managed. Moreover, a significant majority of people prefer forecasts that express uncertainty, while in cases when only a deterministic forecast is provided, people tend to make their inferences about the forecast uncertainty (Morss et al. 2008). This underscores the importance of communicating the forecast uncertainty to the patients. Thus, in this study, we generate and evaluate probabilistic forecasts, rather than focussing on just a point estimate or predefined quantiles. Patient waiting times depend on a multitude of complex *features* (or predictor variables), some of which are also time-varying (such as, staff levels, queue length) and/or patient-specific. Incorporating such individual and time-varying features enables our method to produce personalized forecasts that are updated over time, as new information regarding the ED's utilization and the patient's conditions are observed. The potential nonlinear relationship between high-dimensional ED features and waiting times motivates our decision to adopt a machine learning approach. Specifically, we employ a quantile regression forest (QRF) to estimate conditional probability distributions of the waiting times in a nonlinear and nonparametric framework (Meinshausen 2006). Moreover, to investigate key predictors of ED waiting times, we use a random forest (RF) to derive rankings of feature importance. This information could provide useful insights into patient flow and potential bottlenecks in the ED.

Using a large patient-level dataset, we extract five different categories of features, that quantify: (1) calendar effects (time of arrival), (2) demographics (age, sex), (3) staff levels, (4) ED workload due to patient volumes (attendances), and (5) the severity of patient condition. We provide a comparison of the QRF-based approach with Q-Lasso (Ang et al. 2016), quantile regression (Ding et al. 2010; Sun et al. 2012), *k*-nearest neighbour, and rolling average benchmarks that are typically used in practice. Model evaluation is based on a comparison of distributional, quantile, and point forecast accuracy.

We further discuss three aspects related with practical implementation of the proposed methodology. Firstly, we propose and evaluate a colour-coded (categorical) scheme to effectively communicate forecast uncertainty to the patients. Secondly, we show that our modelling scheme could be used to provide updates of waiting time estimates, by incorporating additional information regarding triage, which becomes available at the time of initial assessment. Finally, we demonstrate that personalized probability distribution estimates of waiting times, when used in conjunction with travel time estimates, could help patients select an ED site from a geographic network of EDs, and could result in more uniform spread of workload among EDs. Patients could access waiting and travel time estimates, on say, a smartphone application. The findings of this study could have direct implications for both patients and emergency service providers.

The paper is arranged as follows. Section 2 describes patient-flow and presents our ED dataset. Section 3 presents features and the quantile regression forest. Empirical results based on a comparison of forecast accuracy are provided in Section 4. Illustrative examples demonstrating the practical applications of the modelling framework are presented in Section 5. Section 6 summarizes the paper and discusses future work.

## 2. ED Patient-flow and Data

### 2.1 Understanding patient-flow in the ED

This study employs data from a major public hospital in the UK. This hospital treats both minor and major injuries, and is operational 24 hours a day. We refer to it as Hospital 1. The policies at Hospital 1 are determined by the NHS, and hence the operations at this hospital are similar to other hospitals in the UK. Figure 1 presents a basic schematic diagram of ED patient flow at Hospital 1. Patients arrive at the ED either via ambulance or any other mode of transport. Upon arrival, patients register at the reception desk, where they provide the following information: name and address, date of birth, reason for visit, name of the general practitioner (doctor) with whom the patient is registered (we denote the time of registration as $t_{reg}$). After registration, patients are requested to take a seat in the ED waiting room until they are called for an initial assessment. At the time of initial assessment (denoted by $t_{assess}$), patients are categorized by a nurse using a: (1) patient group number (code denoting reason for presenting complaint), (2) human resource group (code denoting use of resources), and (3) triage category (to prioritize patients depending upon their severity, the different triage categories are 'minor injury', 'major injury', 'urgent care', or 'resuscitation'). Patients who are triaged as 'minor injury' might need to wait longer in the queue, compared to other patients that arrive later, but who have a more serious health condition. Patients with critical medical needs are triaged as 'urgent care' or 'resuscitation'. These patients are seen with priority by a doctor upon arrival at the ED. Thus, in this study, we only generate waiting time estimates for patients that are triaged as either 'minor injury' or 'major injury' (we refer to these as *low-acuity* patients). Following the initial assessment, a patient returns to the waiting area until they are called by a nurse to start treatment. At the time of treatment (denoted by $t_{treat}$), the patient is seen by a doctor. Depending upon the outcome of the treatment, the patient departs from the ED (patients either leave the hospital, or get admitted to the ICU or some other ward within the hospital).

Studies have suggested that patients are more sensitive to their start-of-treatment time compared to the time-of-departure (Boudreaux et al. 2000). Once with the physician, the patient is typically far more tolerant of the passage of time (Anderson et al. 2007). This motivates us to focus on the time spent waiting until the start of treatment. In addition to receiving, when they arrive, an estimate of their waiting time to treatment, it is helpful to receive updates to this during their wait. We thus update the waiting time estimate for each patient. Specifically, for each low-acuity patient, we model the following two waiting time metrics: (1) $t_1$: time elapsed from registration to start-of-treatment ($t_1 = t_{treat} -$ $t_{reg}$). This prediction is generated at the time of registration. (2) $t_2$: time elapsed from initial assessment to start-of-treatment ($t_2 = t_{treat} - t_{assess}$). This prediction is generated at the time of initial assessment. Note that $t_1$ relates to an initial estimate of $t_{treat}$, and $t_2$ updates this estimate at the time of the initial assessment, whereby information regarding triage and changes in patient-flow are incorporated in the modelling.

**Figure 1.** Schematic diagram of typical patient-flow in the ED.

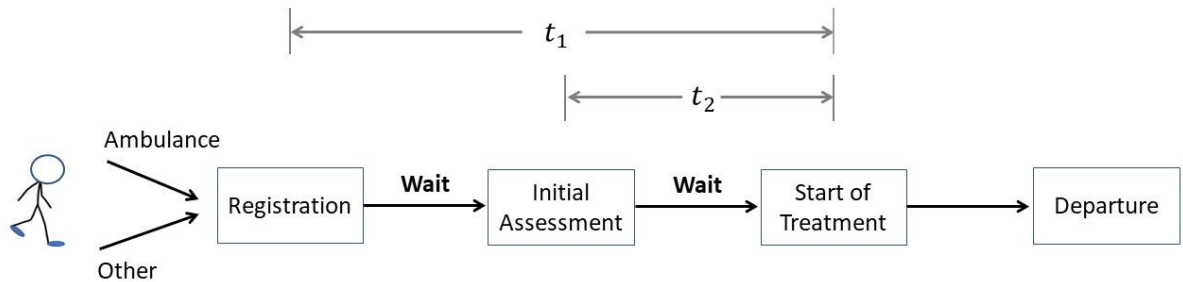

## 2.2 Patient-level ED data

We employ five years of patient-level ED data from 1 January 2014 to 31 December 2018. The data from Hospital 1 is feature-rich. Recording multiple data fields for each patient while working in a high-pressure environment such as an ED, however, also increases the likelihood of having incomplete entries. It is rather unsurprising that EDs are particularly vulnerable to data quality issues (Ward et al. 2015). In practice, domain experts may employ several criteria to deal with data incompleteness. In this study, to avoid any potential bias resulting from data imputation, we adopt stringent criteria for pre-processing where we discard entries with either: incomplete or missing data fields (null entries), faulty timestamps (negative waiting times), or highly unlikely values (waiting times ≥14 hours, age ≥ 110 years). From a total of 454,983 patient-level records, we identified 352,178 patient-level records that were of good quality for the modelling. Out of the 352,178 patients visits: 35.8% were ambulance arrivals, 52.2% were female, 34.2% were admitted to the hospital, the average age was 39 years (standard deviation 27.2 years), 24% were minors (age ≤ 16 years), 25.7% were senior citizens (age ≥ 60 years). Low-acuity patients ($n$ = 334,635) constitute around 95% of the ED attendances, and are the focus of this study. Of these low-acuity patients, those triaged as minor and major injury comprised 43.6% and 51.4% of the total ED visits, respectively.

During the modelling, we first generate a probability distribution estimate of the waiting time from registration to the start-of-treatment ($t_1$) for each of the 334,635 low-acuity patients. We then update this prediction by estimating the waiting time from initial assessment to start-of-treatment ($t_2$) for the 281,910 patients that underwent an initial assessment. Note that, depending upon the severity of the condition and utilization of the ED, some patients (52,725 patients in our dataset) start their treatment without first undergoing an initial assessment. The average start-of-treatment waiting time from point of registration ($t_1$) and from initial assessment ($t_2$) was 86.9 minutes (standard deviation 64.9 minutes) and 72.5 minutes (standard deviation 60 minutes), respectively. We use the first four years (2014–2017) as the in-sample period to estimate model parameters, while the final year (2018) was employed as the out-of-sample period to evaluate forecast accuracy.

**Figure 2.** Plots of attendances, staff count, and waiting times. Panel **2a**: median diurnal attendances, **2b**: median diurnal staff count, **2c**: time series plot for a fortnight of start-of-treatment waiting times from the time of registration ($t_1$), **2d**: autocorrelation function of $t_1$, **2e**: relative frequencies of start-of-treatment waiting times from the time of registration and time of initial assessment ($t_1$ and $t_2$, respectively), and **2f**: median diurnal waiting times ($t_1$ and $t_2$).

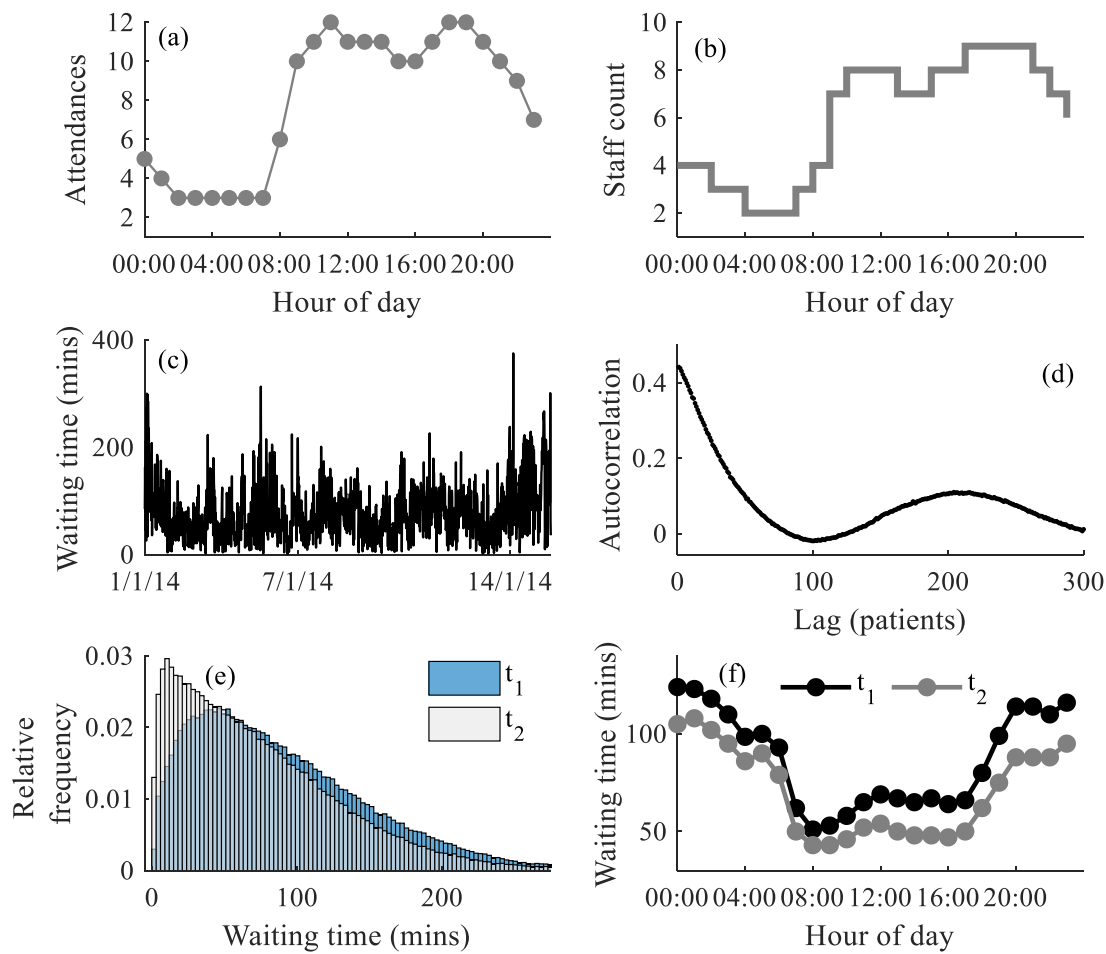

Figure 2 presents median diurnal profiles of *attendances* (i.e. arrivals) and staff count, along with plots of the waiting times ($t_1$ and $t_2$). Attendances at EDs exhibit strong diurnal periodicity, whereby the demand is usually low during the night, and peaks around midmorning and early evening (Figure 2a). A similar diurnal pattern is observed for the median staff count, which is expected, as hospitals allocate more staff to the ED during busier periods of the day (Figure 2b). Figure 2c presents start-of-treatment waiting times ($t_1$) for a two-week period. It is evident that waiting times are highly variable, which underscores the need to generate probabilistic forecasts. Figure 2d presents the autocorrelation plot for waiting times ($t_1$), which is computed at different lags across consecutive patients. This figure shows that waiting times for consecutive patients are autocorrelated, which motivates our decision to employ empirical benchmarks that are based on the rolling average methods used by hospitals in practice. The waiting time distributions are right-skewed, as some patients end up waiting for long hours in the ED (Figure 2e). The diurnal waiting times are typically lower during hours of the day when more staff are available (Figure 2f). Waiting times are overall higher during weekends and Mondays while waiting times on Thursdays are relatively lower. Figure 2 is generated using only the in-sample data.

## 3. Modelling Waiting Times Using a Quantile Regression Forest

In this section, we present our proposed modelling approach. Firstly, we describe the process of feature engineering using the ED data in Section 3.1. The features are used as input variables during the modelling. The quantile regression forest method, which is an extension of random forests, is presented in Section 3.2. Rankings of the most salient features derived from this approach are provided in Section 3.3.

### 3.1 Feature Engineering

Feature engineering is an integral part of modelling, as the accuracy of any statistical method or machine learning approach is conditional upon the quality of input features (Guyon et al. 2008). In the context of this study, feature engineering involves deriving potential predictors of waiting times from the raw patient-level ED data. The patient-level records used in this study are very detailed, which allows us to extract a range of features for each patient. Table 1 presents a list of features extracted from the data along with a brief description. Each feature belongs to one of the following five categories:

(Category 1) **Calendar effects**: These features accommodate periodic variations in waiting times across different periods of the day (diurnal periodicity) and periods of the week (weekly periodicity). Waiting times are typically longer around 8pm-2am (Figure 2f), and during weekends and Mondays (Figure A1). Moreover, EDs often experience anomalous levels of attendances during the holiday periods (Rostami-Tabar and Ziel 2020). To incorporate the effect of anomalous load on waiting times, we use indicator variables to identify holidays (such as Christmas, New Year's Day) and winter proximity days (days around Christmas day).

(Category 2) **Demographics**: These features account for potential differences in waiting times across different demographics (such as age, sex).

(Category 3) **Staff levels**: This feature accounts for the ED service capacity due to staffing. Waiting times are typically low during periods of the day when more staff are available (Figure 2b and 2f). Due to data protection issues, information regarding staff schedules was not directly provided by the hospital. Therefore, we used the unique identifiers/codes of the staff members responsible for discharging a patient to infer the hourly staffing levels.

(Category 4) **ED operations**: These features reflect the state of the ED's operations at any given time, allowing us to represent changes in the ED workload at different points of patient-flow. Relevant features include the numbers of patients in the ED and in different status (e.g., registered but not yet assessed). See Table 1 for the full list of features. For a given low-acuity patient, the waiting time depends on the number of high-acuity patients that are currently in the ED queue. Thus, although we generate waiting time estimates for only the low-acuity patients, we use data for patients across all triage categories for feature engineering to allow for a more complete and accurate representation of the ED workload during the modelling. Moreover, we include features that indicate the number of patients who breached the NHS four- and 12-hour waiting time targets (over the last 24 hours). We also use hourly averages of lagged waiting times as features.

(Category 5) **Patient condition**: These features accommodate the severity of the patient's condition (or presenting complaint) during the modelling. At initial assessment, the following three metrics are used to quantify the patient's condition: (1) Patient group number – indicates the reason for ED episode (e.g., road traffic accident) (2) Human resource group code – indicates the level of resources needed by the patient. (3) Triage category – patients are prioritized for treatment based on their triage level. We are not aware of any other study on predicting patient waiting times based on such detailed information on patient condition. This aspect of modelling leads to a key advantage of our approach – that forecasts can be refined over time as patient information is updated. This will be discussed further in Section 5.1.

**Table 1.** Category, names, and a brief description of different features used for modelling.

| Feature category and name | Brief description |
|---|---|
| Category 1: **Calendar effects** | |
| Hour of day | Arrival hour of day (1 to 24) |
| Hour of week | Arrival hour of week (1 to 24×7) |
| Day of week | Arrival day of week (1 to 7) |
| Month of year | Arrival month of year (1 to 12) |
| Holiday period effects | Indicator variables to accommodate anomalous waiting times during holidays period (0: Normal day; 1: Holiday; 2: Days around Christmas day) |
| Category 2: **Demographics** | |
| Age | Patient age (0 to 110 years) |
| Sex | Patient sex (0: Male, 1: Female) |
| Category 3: **Staffing** | |
| Staff count | Total hourly staff count (inferred via unique staff codes) |
| Category 4: **ED operations** | |
| Total workload | Number of patients in the ED (total, ambulance arrivals, other mode of arrival) |
| Workload from registration to initial assessment | Number of patients in the ED that have registered but have not been assessed (total, ambulance arrivals, other mode of arrival) |
| Workload from initial assessment to start of treatment | Number of patients in the ED that have been assessed but have not started treatment (total, ambulance arrivals, other mode of arrival, number of patients triaged as, minor, major, urgent care, and, resuscitation; and ambulance arrivals triaged as minor, major, urgent care, and, resuscitation) |
| Workload from start of treatment to departure | Number of patients in the ED that have started treatment but have not yet departed (total, ambulance arrivals, other mode of arrival, number of patients triaged as, minor, major, urgent care, and, resuscitation; and ambulance arrivals triaged as minor, major, urgent care, and, resuscitation) |
| 4-hour breach | Number of patients who waited > 4 hours (from registration to departure, over the last 24 hours) |
| 12-hour breach | Number of patients who waited > 12 hours (from registration to departure, over the last 24 hours) |
| Lagged waiting times | Average hourly lagged waiting times for same hour of the day (for 7 previous consecutive days) |
| Category 5: **Patient condition** | |
| Triage level | Category to determine patient's priority for treatment (minor, major, urgent care, and, resuscitation) |
| Human resource group codes | Code to reflect the level of resources needed by the patient (12 alphanumeric codes) |
| Patient group number | Code to identify the reason for ED episode (defined by the NHS as: road traffic accident, assault, deliberate self-harm, sports injury, firework injury, other accident, brought in dead, and other than above) |

## 3.2 Quantile Regression Forests

In this study, the aim of modelling is to estimate the probability distribution function $\hat{F}(y_i|X_i)$ for a target observation $y_i$ that is conditional on the corresponding feature vector $X_i = \{x_{i1}, x_{i2}, \ldots, x_{im}\}$, where $y_i$ denotes the waiting time for the $i^{\text{th}}$ patient, and $X_i$ ($X_i \in \mathbb{R}^m$) is a $m$-dimensional vector of features that quantify properties of the patient, calendar effects, and state of the ED (i.e., features listed in Table 1). For a given patient, $y_i$ refers to $t_1$ at the time of registration, while at the time of initial assessment, $y_i$ refers to $t_2$. Given $n$ low-acuity patients having waiting times, $Y = \{y_1, y_2, \ldots, y_n\}$ (the *label vector* with size $n \times 1$), and corresponding features represented by $X = \{X_1, X_2, \ldots, X_n\}$ (the *feature matrix* with size $n \times m$), we aim to train a model (denoted by $\Omega$, with parameter matrix $B$), which is a mapping from the input features to the corresponding waiting time distribution, $\hat{F}(y_i|X_i) = \Omega(X_i, B_i)$.

QRF is a generalization of the popular random forest (RF) method. RF is an ensemble machine learning method that has commonly been used to generate accurate predictions using high-dimensional features (Breiman 2001). The performance of RF has been shown to be robust under the presence of noisy or highly correlated features (Breiman 2001). Besides prediction, RF can be used to derive rankings of feature importance which can help make valuable inference from the data (Hastie et al. 2009). While RF provides an accurate approximation of the conditional mean of the target variable in a nonlinear and nonparametric framework, QRF estimates the conditional probability distribution of the target variable (Meinshausen 2006).

To construct a QRF, we grow a large set of regression trees. While growing each tree and node, randomness is incorporated during the selection of features. For a given tree, a bagged version of the training data is used. For each node, a random subset of features is used for splitpoint selection while approximating the target variable. A tree is grown by splitting of the bootstrap training sample such that it minimizes the total impurity (sum of squared deviations about a group mean). The process of splitting continues until a minimum leaf size has been achieved. Given the set of trees, dropping a new data point down each tree reaches a leaf node that produces a single forecast (observation) of the target variable. While RF estimates the conditional mean of the target variable by averaging such observations over the set of trees, QRF stores the value of all observations in the leaf nodes to estimate an empirical cumulative distribution function of the target variable.

For a single tree, denoted by say $T(\theta)$, which is grown using a random feature subset $\theta$, the point forecast of the mean obtained using a new feature vector $X_{new}$ is computed as the average of the subset of target values $y_i$ in the training sample associated with feature vectors $X_i \in l(X_{new}, \theta)$, where $l(X_{new}, \theta)$ is the leaf node that contains $X_{new}$. Mathematically, the forecast of the mean is given by $\sum_{i=1}^{n} w_i(X_{new}, \theta) y_i$, where the weights $w_i(X_{new}, \theta)$ are given by:

$$w_i(X_{new}, \theta) = \frac{1_{\{X_i \in l(X_{new}, \theta)\}}}{\left|j : X_{j \in l(x,\theta)} \in l(X_{new}, \theta)\right|}.$$

For a forest of $n_{tree}$ regression trees, the weights from each tree are averaged as: $w_i(X_{new}) = 1/n_{tree} \left( \sum_{j=1}^{n_{tree}} w_i(X_{new}, \theta_j) \right)$, where $\theta_j$ denotes the feature subset used for growing the $j^{th}$ tree. In RF, the conditional mean $E(y|X = X_{new})$ can be estimated as $\sum_{i=1}^{n} w_i(X_{new}) y_i$. QRF provides a natural extension for probabilistic forecasting by estimating the conditional distribution as:

$$\hat{F}(y|X = X_{new}) = \sum_{i=1}^{n} w_i(X_{new}) I(y_i < y).$$

Instead of computing the average of observations in leaf nodes as done in RF, QRF keeps track of all observations and their weights, for all leaves and across all trees, to estimate the distribution function of the target variable. This estimate of the conditional distribution function is asymptotically consistent (Meinshausen 2006). In a recent study, using features extracted from flight- and passenger-level data, regression trees have been shown to be effective in predicting the probability distributions of connection times and number of arrivals at an airport (Guo et al. 2018). In our study, 500 regression trees are grown (Breiman 2001), and one-third of the total features are used for splitpoint selection at each node (Meinshausen 2006), while nodes with less than 5 observations were not split any further.

### 3.3 Rankings of Feature Importance

We also use the RF underlying the QRF approach to identify the most salient features associated with predicting waiting times at different stages of patient-flow in the ED (see, for example, Breiman 2001). Teasing out the set of features with the strongest predictive power can help service providers focus their attention on a small set of factors that could influence patient flow, and inform further identification studies to uncover causal relationships. The process of finding the most salient subset of features (which typically involves removing irrelevant, redundant, and noisy features) is referred to as *feature selection*, and has been an area of active research (Guyon et al. 2008).

Figure 3 provides the feature importance scores for the 20 most salient features obtained from random forest (Brieman 2001), as described in Section 3.2, using only the in-sample data. For both $t_1$ and $t_2$, age was the single most salient feature. This can perhaps be explained by the fact that Hospital 1 has a dedicated children's emergency ward that prioritizes and treats only minors (age $\leq$ 16 years). This finding was further corroborated by an analysis of the data which revealed that approximately a quarter of the total attendances were minors for whom the mean waiting times were considerably shorter than the remaining low-acuity patients (77 minutes vs 90 minutes, for $t_1$), moreover, the two waiting time distributions were significantly different (using a Kolmogorov-Smirnov test with 1% significance level). Interestingly, the bulk of the most salient features are associated with ED operational characteristics, particularly workload due to patient volumes and calendar effects.

**Figure 3.** Feature importance scores for modelling: (a) $t_1 -$ registration to start-of-treatment, and (b) $t_2 -$ initial assessment to start-of-treatment. *Note*: a higher score denotes a higher-ranked feature.

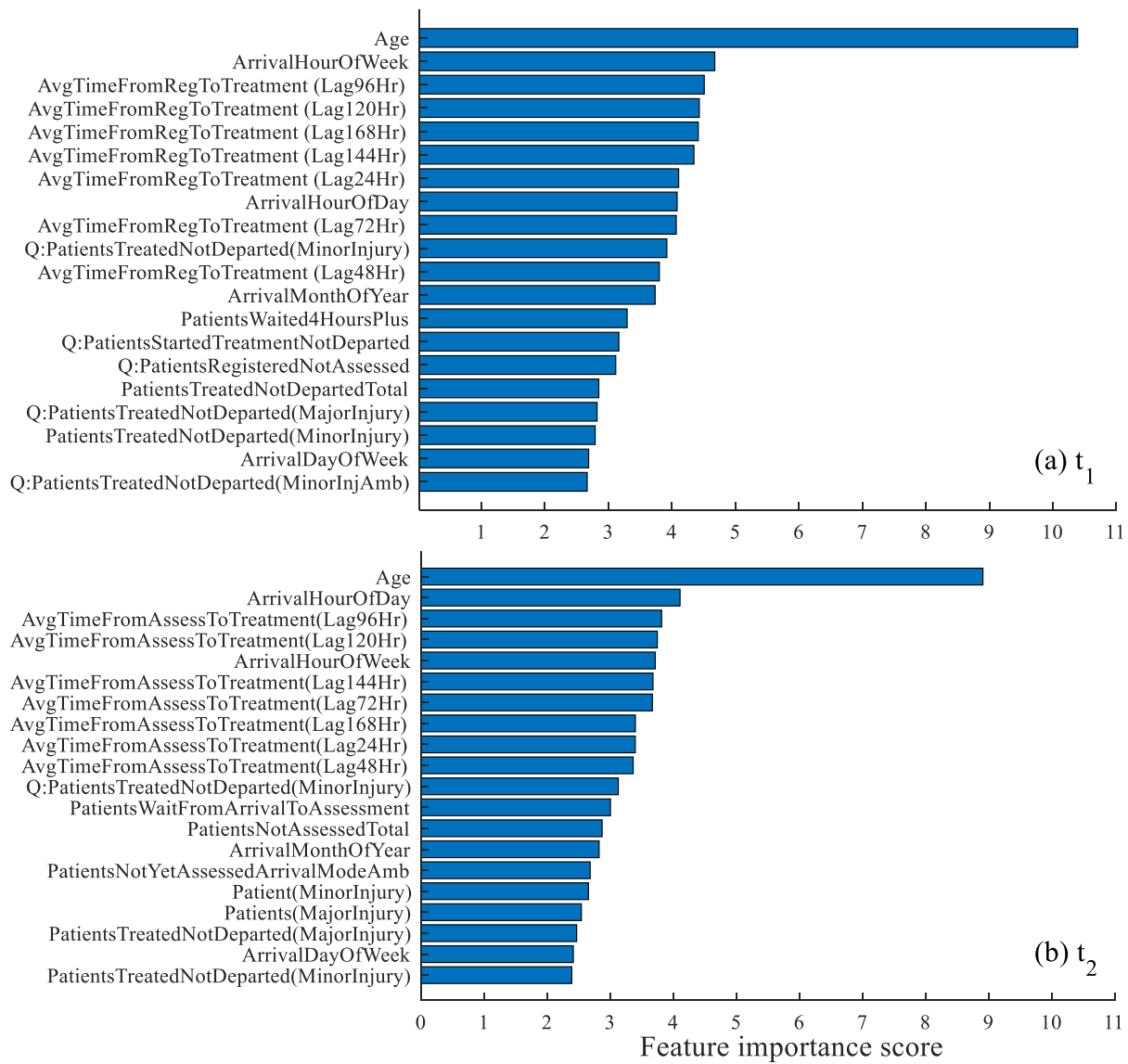

Features based on historical rolling averages also ranked higher, which indicates the presence of correlations at multiple lags in waiting times (as evident from Figure 2d). Note that features with prefix Q in the figure denote features that were inspired by the fluid model estimators of Ang et al. (2016). Note that staff count did not show up among the list of most salient features presented in Figure 3. Since both attendances and staff count exhibit a strong diurnal periodicity, it is possible that the diurnal variation in staff counts is largely captured by features that quantify attendances (i.e. ED workload). We caution that, in the absence of proper identification, this observation does not imply the absence or weakness of causal effects between staffing and waiting times.

## 4. Evaluating Out-of-sample Forecasting Accuracy

In this section, we compare the performances of QRF and a set of benchmarks in terms of their distributional, quantile, and point forecast accuracy. Model rankings are assessed based on the accuracy of forecasting $t_1$ and $t_2$, using data for the one-year out-of-sample period (2018). Using our more detailed ED dataset, we were able to extract a richer set of features compared to previous studies. For example, to forecast patient waiting times, Sun et al. (2012) included information regarding the date, time of triage, time of consultation, and patient acuity category as features for a quantile regression model. To ensure a fair evaluation of different forecasting strategies, we decided to employ all features (as shown in Table 1) as inputs, across all methods. Forecasts for $t_1$ were generated using features from the first four categories. For forecasting $t_2$, features from all five categories were used during the modelling.

### 4.1 Benchmark Methods

As naïve benchmarks, we present three empirical methods in Section 4.1.1. As sophisticated benchmarks, we present quantile regression (Ding et al. 2010; Sun et al. 2012), Q-Lasso (Ang et al. 2016), and *k*-nearest neighbour, in Sections 4.1.2-4.1.4. For parameter estimation, we used the last year of the in-sample data (2017) as the cross-validation hold-out sample.

### 4.1.1 Empirical Methods

The empirical methods, although simplistic in their mathematical formulation, are associated with low computational complexity, which makes them attractive for deployment in the ED. We use the following three empirical methods (based on rolling averages) for forecasting waiting times:

(1) **Empirical 4-hours** – empirical distribution of waiting times observed in the previous four-hours. For EDs that publish waiting time estimates in the US, a four-hour rolling average has become the conventional choice (Dong et al. 2019).

(2) **Empirical $p$-hours** – waiting times for the last $p$ consecutive hours ($h$-1, $h$-2, …, $h$-$p$), where $h$ denotes the hour of arrival for the current patient.

(3) **Empirical $q$-periods** – waiting times of the previous $q$ periods conditional on the same period of the day ($h$-24, $h$-2×24, …, $h$-$q$×24).

To produce these three estimates of the probability distribution of $t_1$, we compute averages of historical waiting times from the time of registration to start-of-treatment. For $t_2$, we employ waiting times from the time of initial assessment to start-of-treatment. We estimate the optimal values of $p$ and $q$ based on the minimization of the sum-of-squared errors over the cross-validation hold-out sample.

### 4.1.2 Quantile Regression

Least squares regression has been employed for forecasting waiting times (Asaro et al. 2007). However, least squares regression only provides an estimate of the conditional mean, while ignoring other aspects of the conditional distribution. This is a serious limitation, as waiting time distributions are heavily right-skewed (Sun et al. 2012), as evident from Figure 2e. Quantile regression has thus been proposed for modelling waiting times (Ding et al. 2010; Sun et al. 2012), as it allows for a more detailed characterization of the ED data by quantifying the impact of features on the distribution of waiting times. The quantile regression estimator, for the $\tau$ quantile, is based on the minimization of an asymmetric absolute loss function (see, for example, Koenker and Bassett 1978). In this study, we estimate quantile regression models for the following quantiles $\tau$ = {5%, 15%, 25%, 35%, 45%, 50%, 55%, 65%, 75%, 85%, 95%}. To construct distributional forecasts using quantile regression, we linearly interpolate between the estimated quantiles and treat the minimum and maximum of the in-sample data as bounds of the forecast distribution.

### 4.1.3 Q-Least Absolute Shrinkage and Selection Operator (Q-Lasso)

This method was proposed by Ang et al. (2016) for generating point estimates of patient waiting times. Q-Lasso combines concepts from queuing theory with the statistical modelling approach of Lasso, which employs the following objective for model estimation:

$$min_\beta \frac{1}{n}\sum_{i=1}^{n}(y_i - \beta^T X_i)^2 + \lambda ||\beta||_1$$

where $||\beta||_1 = \sum_{j=1}^{m}|\beta_j|$ is the L1 norm of the coefficients $\beta$ ($\beta \in \mathbb{R}^m$), and $\lambda$ is the regularization parameter. The penalty term $\lambda||\beta||_1$ helps prevent model over-fitting by forcing coefficients of less salient features to go to zero. Lasso thus makes the regression model more parsimonious. For $\lambda$ equal to 0, Lasso defaults to ordinary least squares. This method predicts waiting times as a linear function of features within a parametric modelling framework. We estimate $\lambda$ using 5-fold cross-validation (for details, see Hastie et al. 2009). We used this method to generate only point forecasts as there are challenges associated with Lasso while estimating standard errors (Goeman 2010), and this challenge is further exacerbated by the fact that waiting times predictions intervals cannot take negative values.

The Q-Lasso method incorporates fluid model estimators as candidate predictors in Lasso. In the context of modelling patient waiting times, these estimators generalize to the ratio: *workload/processing rate*. Workload is associated with the number of patients that must be seen before a new low-acuity patient can start treatment (Category 4 features), whereas the processing rate depends on the number of available staff (Category 3 features). Following Ang et al. (2016), for a given workload feature (say, total attendances), an additional feature was included in the model, which was calculated by dividing that feature with the corresponding hourly staff count (e.g. total attendances/staff count).

### 4.1.4 $k$-Nearest Neighbour ($k$-NN)

Although it has not been used before in the literature for forecasting waiting times, we use $k$-NN as it can be interpreted as a more sophisticated adaptation of the rolling average methods that are used in practice. While rolling average methods rely on averaging the observations of waiting times (for say, the last $p$-hours, or $q$-periods), $k$-NN generates waiting time estimates for a given patient based on the waiting times of the previous $k$ most similar patients. Similarity, between any two patients is quantified by Euclidean distance in the feature space. As the five categories of features constitute a different number of features (Table 1), Euclidean distance was computed separately for each feature category, to ensure that each feature category received equal weight during the modelling. Features were standardized to have zero mean and unit standard deviation. For a given patient, probabilistic forecasts were generated by using the empirical distribution of the waiting times for the historical $k$ most similar patients. We estimate $k$ by minimizing the sum-of-squared errors over the cross-validation hold-out-sample.

## 4.2   Evaluating Distributional Forecasts

Figure 4 provides an illustration of distributional forecasts generated by QRF for two days chosen at random from the out-of-sample period. The forecast origin was midnight at the start of 8 April 2018 (Sunday). The median of the forecast distribution was issued as the point forecast. The point forecast and its uncertainty exhibit a similar diurnal pattern to those shown in Figure 2f. Encouragingly, in Figure 4, the 90% interval of the forecast distribution encompasses the vast majority of the actual observations.

**Figure 4.** Summaries of out-of-sample distributional forecasts for: (a) $t_1$ − registration to start-of-treatment, and (b) $t_2$ − initial assessment to start-of-treatment. *Note*: the shaded regions correspond to the different quantile ranges of the forecast distribution (encompassing 90% and 50% area of the forecast distribution, centered around the median).

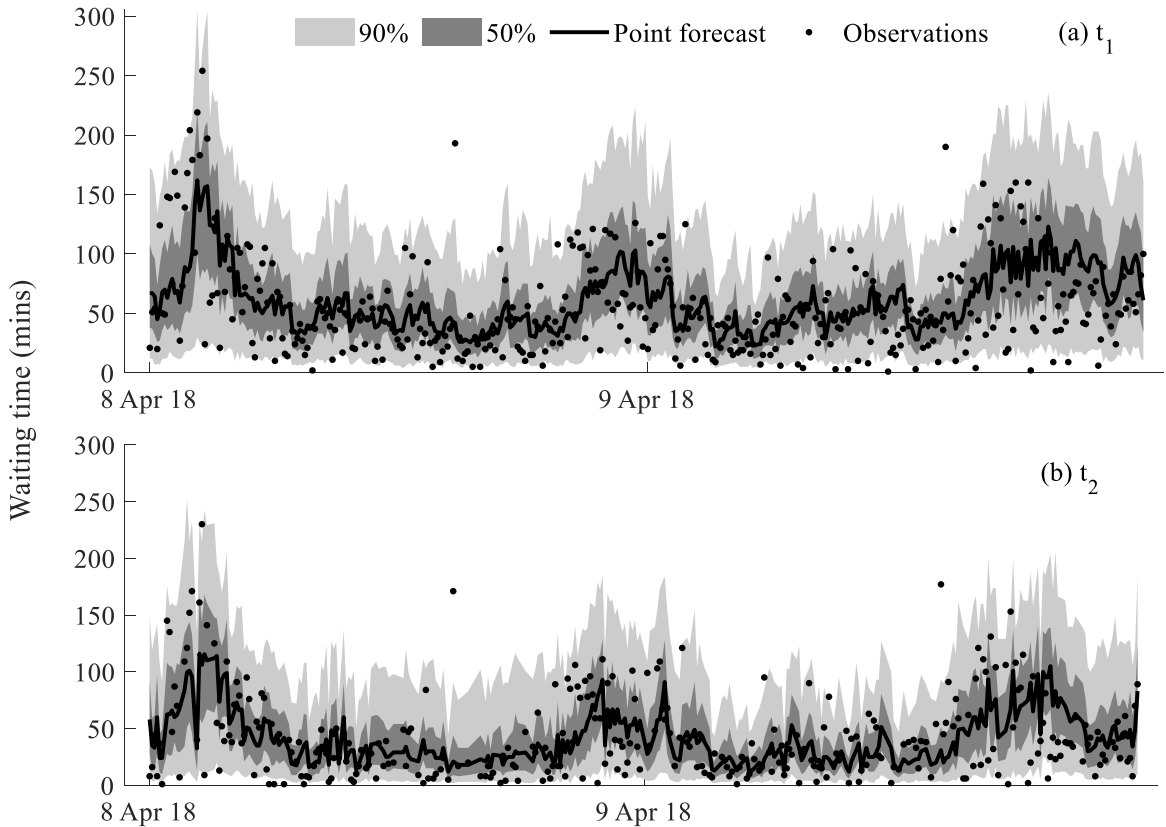

To quantify distributional forecast accuracy, we use the continuous ranked probability score (CRPS), which is a strictly proper scoring rule that quantifies both *sharpness* (concentration or peakedness of the forecast distribution) and *calibration* (statistical consistency between forecast distribution and actual observations). We use the expectations form of the CRPS (Gneiting and Raftery 2007), defined as:

$$CRPS = E_F|Y - y| - \frac{1}{2} E_F|Y - Y'|$$

where $Y$ and $Y'$ are independent samples drawn from the forecast probability distribution function (we draw 1000 samples), each having the same underlying distribution $F$, $E_F$ is the expectation with respect to the distribution $F$, and $y$ is the corresponding actual waiting time.

In Table 2, we present the CRPS values averaged across all patients in the out-of-sample period. QRF is the most accurate method for forecasting both $t_1$ and $t_2$. QRF provides a reduction of more than 20% in the CRPS compared to the empirical 4-hour (typically used in practice), and more than 10% reduction in comparison to the best empirical method (empirical $p$-hours). QRF also outperformed the other benchmark methods, though the improvement over quantile regression (the best performing benchmark) was relatively modest. Surprisingly, $k$-NN was outperformed by the empirical $p$-hours method. As explained earlier, we were not able to produce probabilistic forecasts from Q-Lasso.

**Table 2.** Mean CRPS for distribution forecasting of: (a) $t_1$ − registration to start-of-treatment, and (b) $t_2$ − initial assessment to start-of-treatment.

| Forecasting Method | $t_1$ | $t_2$ |
| --- | --- | --- |
| Empirical 4-hours | 31.7 | 29.8 |
| Empirical $p$-hours | 28.0 | 26.3 |
| Empirical $q$-periods | 34.4 | 34.2 |
| Quantile regression | 25.5 | 24.3 |
| $k$-nearest neighbour | 31.3 | 29.2 |
| Quantile regression forests | **24.8** | **22.4** |

*Note*: lower CRPS values are better (lowest values are highlighted in **bold**).

In Table 2, the errors associated with $t_2$ are overall lower than the corresponding errors for $t_1$. This shows that the modelling approaches are more accurate as one gets closer to the start-of-treatment time. This is due to three reasons: (1) for any given patient, $t_2$ is lower in magnitude than $t_1$ by construction, (2) features that quantify ED workload due to patient workflow are updated at the time of assessment, and (3) crucially, additional features (that quantify the severity of the patient's condition) are incorporated while modelling $t_2$. Encouragingly, compared to the CRPS for $t_1$, QRF provides the highest percentage reduction (9.7%) in the CRPS values while estimating $t_2$. These results highlight that compared to linear modelling approaches, QRF is better equipped to take advantage of information (feature) updates, as the potential relationship between such new information (features) and the target variable can be nonlinear. For $t_1$, Figure 5 presents the out-of-sample CRPS values for each patient plotted against their corresponding actual waiting times. The magnitude of probabilistic forecast error is notably larger for patients that wait for exceedingly long hours in the ED. The accuracy is best for patients for whom the waiting time was close to the median in-sample waiting time of 73 minutes.

**Figure 5.** CRPS values for all patients in the out-of-sample period plotted against their corresponding actual waiting time ($t_1$, in minutes). The left and bottom panel figures present the probability distributions of the CRPS and waiting times, respectively.

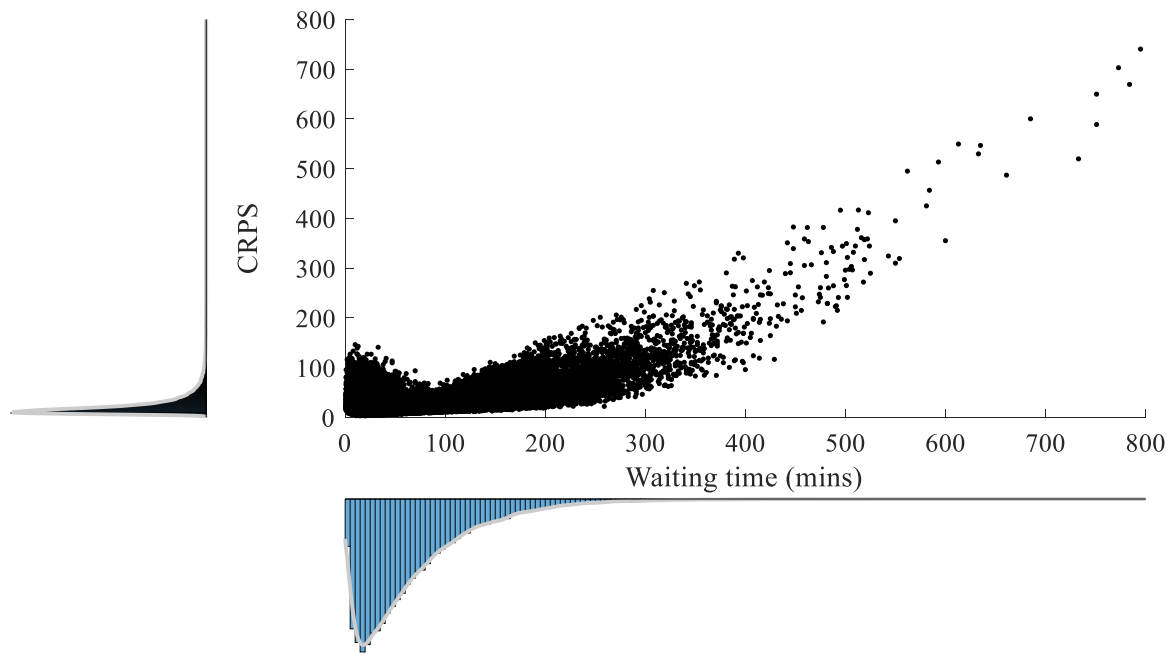

### 4.3 Evaluating Categorical Forecasts

The effective and concise communication of forecast uncertainty is a challenging task, since visualising probability distributions is not necessarily an intuitive task for patients without statistical backgrounds. To address this, we propose the following interpretable colour-coded reporting scheme: *Green* (for *low* waiting times ≤ 45 minutes), *Amber* (for *medium* waiting times: 45 minutes < waiting times ≤ 120 minutes), and *Red* (for *high* waiting times > 120 minutes). For example, a forecast issued as: Green (10%), Amber (70%), Red (20%); would imply a 10% probability that the patient would start treatment within 45 minutes, and so on. We employ this visual form of forecast representation as people are typically familiar with colour-coded information (e.g., traffic signals, air quality indices, etc.).

Converting the continuous target variable (waiting times) into a categorical variable (green, amber, red) transforms the modelling challenge from a regression problem into a classification task. Specifically, the aim of this classification task is to predict the discrete probabilities of the three categories of waiting times. We do this using the probability distributions of the different methods compared in Section 4.2, as well as two additional methods: a three-class classifier and a multiple binary classifier. For multi-class and binary classification, we use a random forecast classifier. Converting the waiting times into a categorical variable resulted in imbalanced data, i.e. a different number of training observations belonging to the three classes. Data imbalance makes the learner more prone to over-classify the majority class. To tackle the issue of an imbalanced data during classification, we assign a larger weight to the underrepresented class, whereby the class weights sum to one (see, for example, He and Garcia 2009). To evaluate categorical forecasts, we use the ranked probability score (RPS), which is the score for discrete distributions that is analogous to the CRPS (Epstein 1969). Table 3 presents the average out-of-sample RPS. For regression-based methods (Table 2), we estimate discrete probabilities associated with the three classes from the corresponding forecast distribution. Encouragingly, QRF outperformed other modelling schemes in the classification task, reporting better RPS. QRF could be thus be used for estimating discrete probabilities for the colour-coded waiting time categories.

**Table 3.** RPS (×100) for categorical forecasting of: (a) $t_1$ − registration to start-of-treatment, and (b) $t_2$ − initial assessment to start-of-treatment.

| Forecasting Method | $t_1$ | $t_2$ |
|---|---|---|
| Empirical 4-hours | 19.5 | 18.5 |
| Empirical $p$-hours | 17.2 | 16.2 |
| Empirical $q$-periods | 21.5 | 21.1 |
| Quantile regression | 15.7 | 14.4 |
| $k$-NN | 19.6 | 18.3 |
| Quantile regression forests | **15.2** | **13.6** |
| Three-class classifier | 15.6 | 14.0 |
| Multiple binary classifiers | 15.6 | 14.0 |

*Note*: Lower RPS values are better (lowest values are highlighted in **bold**).

## 4.3 Evaluating Quantile Forecasts

To evaluate quantile forecasts, we use *unconditional coverage*, which measures the percentage of observations that are lower than the $\tau$ quantile forecast. Ideally, this percentage should be $\tau$. Figure 6 presents the unconditional coverage for $t_1$ and $t_2$, averaged across all low-acuity patients, for $\tau = 5\%$, $15\%$, $25\%$, $35\%$, $45\%$, $50\%$, $55\%$, $65\%$, $75\%$, $85\%$ and $95\%$. In Figure 6, values closer to the diagonal line (ideal coverage) are better. It can be seen from the figure that the unconditional coverage is rather impressive for empirical ($p$-hours), quantile regression (QREG), and QRF. The performances of $k$-NN and the two empirical benchmarks (based on a 4-hour window and $q$-periods) are much poorer, which is consistent with the rankings obtained using the CRPS and RPS.

**Figure 6.** Unconditional coverage for quantile forecasts of waiting times for: (a) $t_1$ − registration to start-of-treatment, and (b) $t_2$ − initial assessment to start-of-treatment. *Note*: values closer to the diagonal are better.

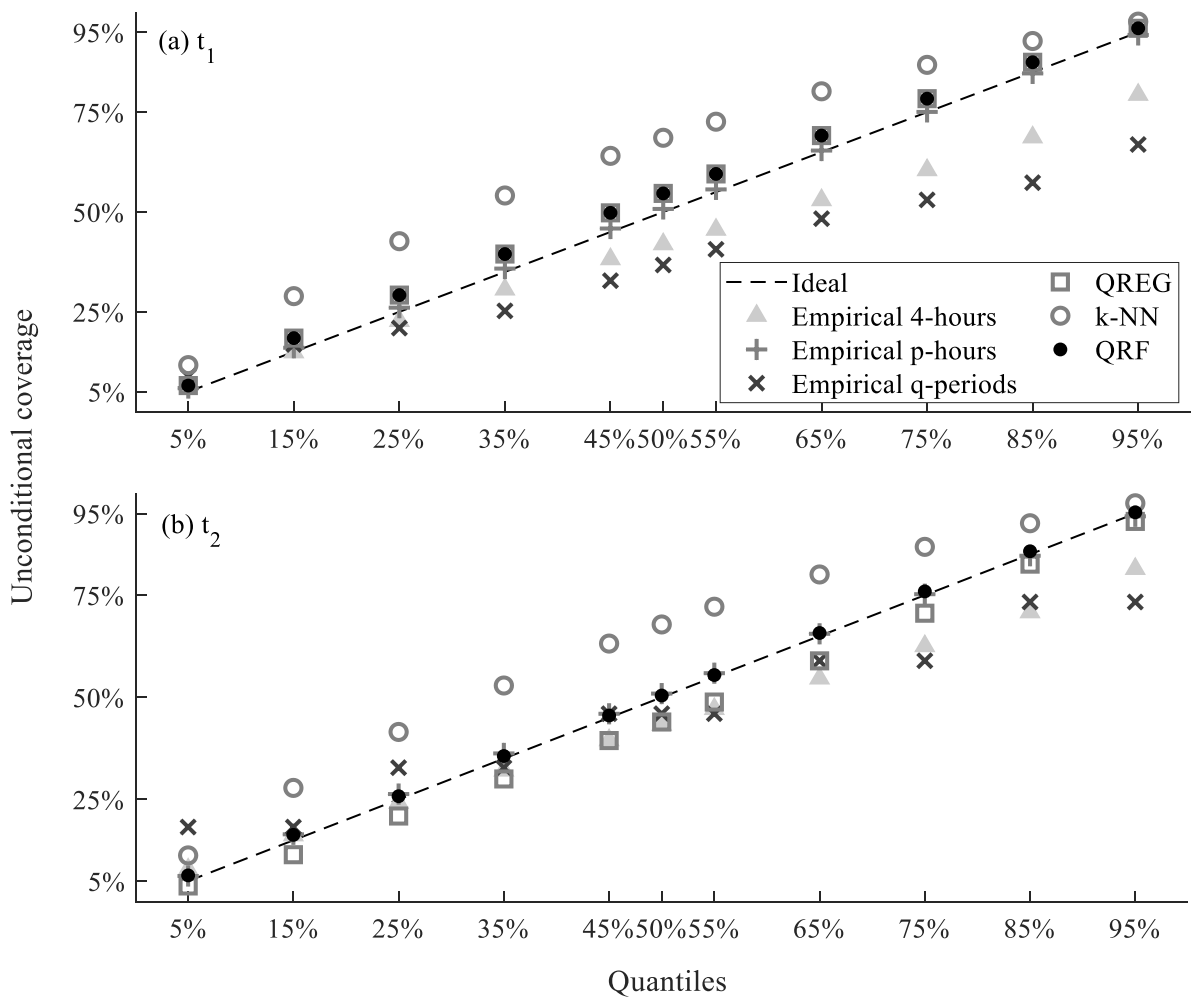

## 4.4 Evaluating Point Forecasts

In Figure 7, we plot the median of the out-of-sample distributional forecasts of QRF (for $t_1$ and $t_2$) across different hours of the day using the one-year out-of-sample period. This figure shows that QRF is overall able to accommodate the diurnal periodicity in waiting times. For evaluating point forecasts, we use the root mean square error (RMSE) and the mean absolute error (MAE). For a quadratic loss function, the mean is the optimal forecast, whereas the median of the forecast probability distribution is the optimal forecast if the loss function is symmetric piecewise linear (Gneiting 2011). In view of this, we issue the mean of each forecast distribution as a point forecast, for evaluation based on the RMSE. For evaluation based on the MAE, we issue the median of each forecast distribution as a point forecast. Forecast of the mean estimated using the Q-Lasso is used as a point forecast.

**Figure 7.** Median diurnal variations in out-of-sample actual and predicted waiting times for: (a) $t_1$ − registration to start-of-treatment, and (b) $t_2$ − initial assessment to start-of-treatment.

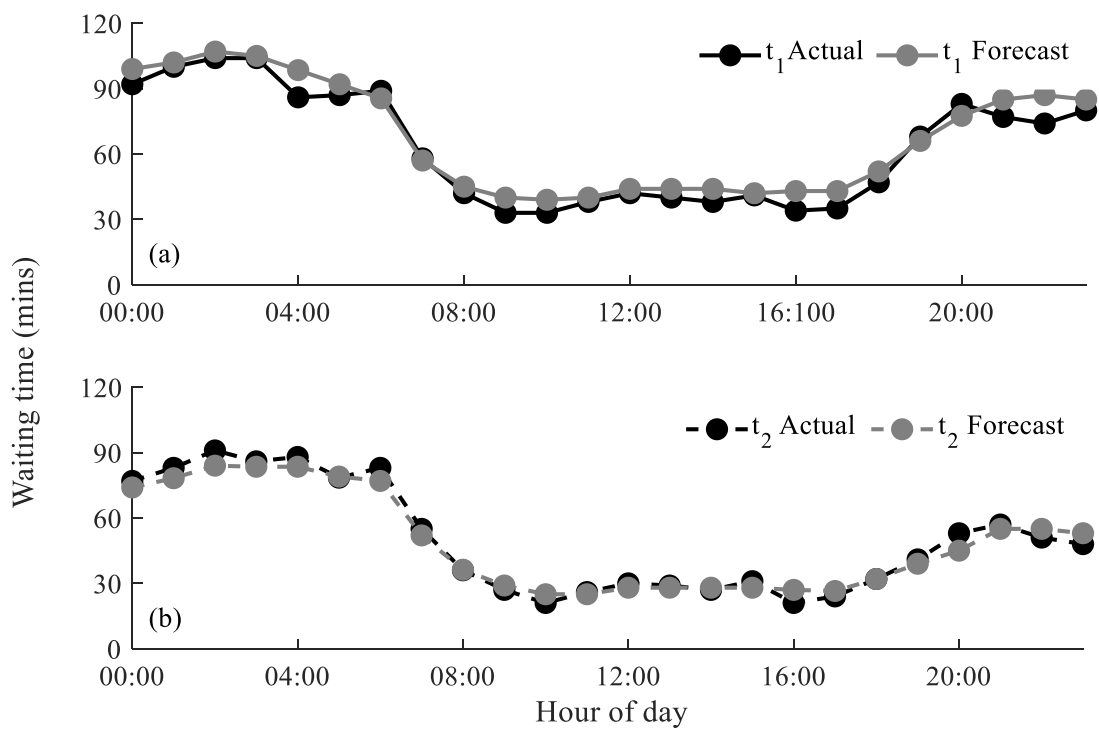

In Table 4, we present the RMSE and MAE results. For point forecasting, QRF is the best performing method. The performances of Q-Lasso and quantile regression are rather similar, whereby both models outperform the empirical benchmarks. The performance of $k$-NN is not competitive with the best empirical benchmark.

**Table 4.** RMSE and MAE for point forecasting of: (a) $t_1$ – registration to start-of-treatment, and (b) $t_2$ – initial assessment to start-of-treatment.

|  | RMSE | | MAE | |
|---|---|---|---|---|
| **Forecasting Method** | $t_1$ | $t_2$ | $t_1$ | $t_2$ |
| Empirical 4-hours | 60.4 | 57.5 | 42.5 | 40.2 |
| Empirical $p$-hours | 55.9 | 53.3 | 39.8 | 37.2 |
| Empirical $q$-periods | 63.4 | 62.3 | 43.9 | 44.3 |
| Quantile regression | 50.8 | 49.0 | 36.0 | 33.9 |
| Q-Lasso | 51.0 | 48.6 | 37.1 | 33.0 |
| $k$-nearest neighbour | 60.7 | 57.4 | 45.1 | 41.9 |
| Quantile regression forests | **50.1** | **46.6** | **35.1** | **31.5** |

*Note*: lower RMSE and MAE values are better (lowest values are highlighted in **bold**).

## 5. Implementation

In this section, we discuss aspects related to the practical implementation of the proposed probabilistic modelling framework. Section 5.1 focusses on publishing and updating waiting time estimates at different stages of patient-flow in the ED. Section 5.2 demonstrates that estimates of waiting times, when used in conjunction with travel times, could help patients make informed decisions while selecting an ED site from a network of neighbouring EDs.

### 5.1 Implementing the Forecasting Scheme at the ED with Information Updates

We demonstrate how the proposed forecasting scheme could be used in practice to communicate personalized waiting times, which patients could access using say a smartphone, as they wait in the ED. For demonstration, we select data for an actual patient from the out-of-sample period, and for ease of explanation, we refer to this patient as John. Figure 8 demonstrates the patient-flow for John at the ED, who registered at 19:39 on 9 April 2018 (non-ambulance arrival). Using features for calendar effects (time and day of arrival etc.), demographics (90 year old male), staff count, and state of the ED patient work-load at John's time of registration (number of total patients, ambulance arrivals, etc), we generate the forecasts for the waiting times.

**Figure 8.** Schematic diagram illustrating ED patient-flow and forecasts. Panel **A** presents the actual timestamps of registration, initial assessment, and treatment (for a 90 year old male, triaged as 'major injury', patient group '80' and code 'VB04Z' during the initial assessment). Panel **B** and Panel **C** shows the point forecast, probabilistic forecast and categorical probability forecast, for $t_1$ and $t_2$, respectively. Forecasts shown in Panel B and Panel C are generated at the time of registration ($t_{reg}$) and time of initial assessment ($t_{assess}$), respectively.

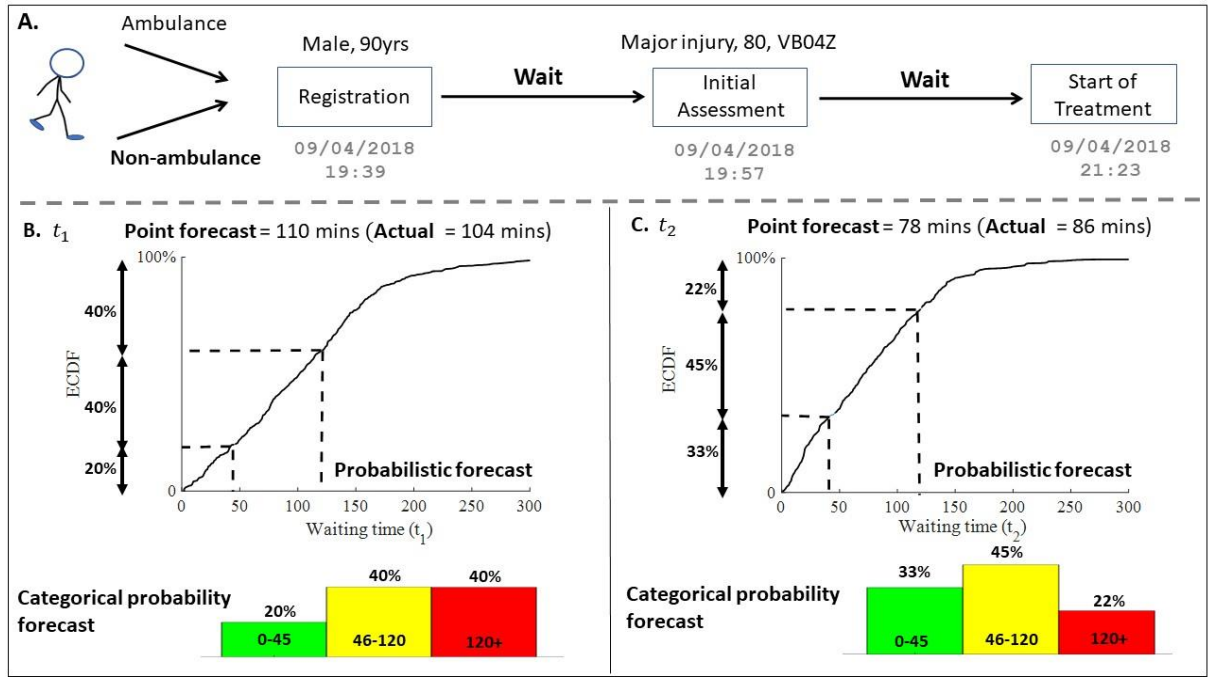

The point forecast from QRF indicated that John would start treatment in 110 minutes from the time of registration (actual $t_1$ was 104 minutes). The colour-coded scheme for categorical forecasts conveyed a 20% chance of waiting less than 45 minutes, and a likelihood of 40% for waiting between 46 to 120 minutes, and a 40% probability of waiting longer than 120 minutes, as evident from the empirical cumulative distribution function (ECDF) of the waiting time forecasts. Note that the estimate for $t_1$ was generated at the time of registration. At 19:57, John underwent initial assessment, and at this point he was triaged as 'major injury' and assigned a patient group number '80' (code denoting reason for presenting complaint) and a code 'VB04Z' (code denoting use of resources). This additional triage related information was incorporated in the model at the time of initial assessment. Moreover, since the state of patient-flow at the ED can change rather quickly, we update feature values for the ED workload (Category 4 features) to accommodate changes in the ED from 19:39 to 19:57. The time spent by John from registration to the start-of-treatment was also fed as a feature in the model for modelling $t_2$. The updated point forecast from QRF indicated that John would start treatment in 79 minutes from the time of initial assessment (actual $t_2$ was 86 minutes). The updated colour-coded scheme now reflected a 33% probability of starting treatment within the next 45 minutes, a 45% probability of starting treatment between 46 to 120 minutes, and only a 22% probability of starting treatment after two hours. Note the initial forecast was generated at the time of registration, and then updated at the time of initial assessment. This illustration presents a possibility in which EDs could convey waiting time estimates (and associated uncertainty) to each low-acuity patient.

## 5.2 Implementing the Forecasting Scheme Remotely: A Demonstration

Studies have emphasized the importance of informed routing decisions to get the right patient to the right provider at the right time (Singh et al. 2020), as access to waiting times for different healthcare providers can assist patients and first responders to make better decisions in selecting a hospital site, which in turn has been shown to reduce actual wait times (Xie and Youash 2011). Estimates of waiting times, when used in conjunction with travel times, could potentially be used for ambulance routing and diversion, to facilitate the uniform spread of load within a network of neighbouring hospitals (Deo and Gurvich 2011; Xu and Chan 2016). To assist patients in selecting an ED from a network of EDs with different waiting times, smartphone applications have been proposed (such as, Waitless, NHS Quicker). However, the aforementioned studies and smartphone applications do not take into account the uncertainty in travel and waiting times.

We show that decision-makers (such as patients, first responders) could select an ED from a network of EDs based on probabilistic modelling of travel and waiting times. This analysis necessitates data for more than one ED. We include additional data from another public hospital, situated in the same region as Hospital 1. We refer to this hospital as Hospital 2. Specifically, we use five years of data from 1 January 2014 to 31 December 2018 from Hospital 2, comprising 180,715 patient-level records. As done earlier, data for the first four years was used for training, while data from the final year was used for evaluation. Encouragingly, out-of-sample model comparison using data from Hospital 2 revealed that QRF generated the most accurate forecasts of the benchmark methods that we considered in Section 4, consistent with our findings with Hospital 1.

We simulate patient choices while selecting an ED site by accommodating uncertainty in travel and waiting time estimates. Due to data protection issues, we did not have access to patient locations. Thus, for demonstration purposes, we randomly assign patients to postcodes from the geographic neighbourhood of the hospital sites. To estimate travel times, using each patient's timestamps (time of selecting an ED) and postcode, we accessed the following data from Google maps for both hospitals: distance, minimum driving time, and maximum driving time. Studies have investigated a range of travel time distributions under different traffic conditions (Guessous et al. 2014). For simplicity, we assumed the travel times to follow an exponential distribution in this study. The minimum travel time was treated as the true lower bound, and we consider maximum travel time as the 99% quantile of the exponential distribution. We simulate data for a day in the out-of-sample period for both hospitals ($n = 318$ patients).

**Figure 9.** Schematic diagram illustrates patient decision-making regarding the choice between Hospital 1 and Hospital 2. Panel **A** presents the demographics, time of decision, and location. Panel **B** depicts travel routes to the two EDs. Panel **C** shows the exponential distribution of travel times. Panel **D** shows that for each sample of travel time, we have a corresponding estimate for the time of registration and feature vector. Panel **E** depicts a feature vector being used as an input for a regression tree. Panel **F** shows the empirical cumulative distribution function (ECDF) for combined travel and waiting times.

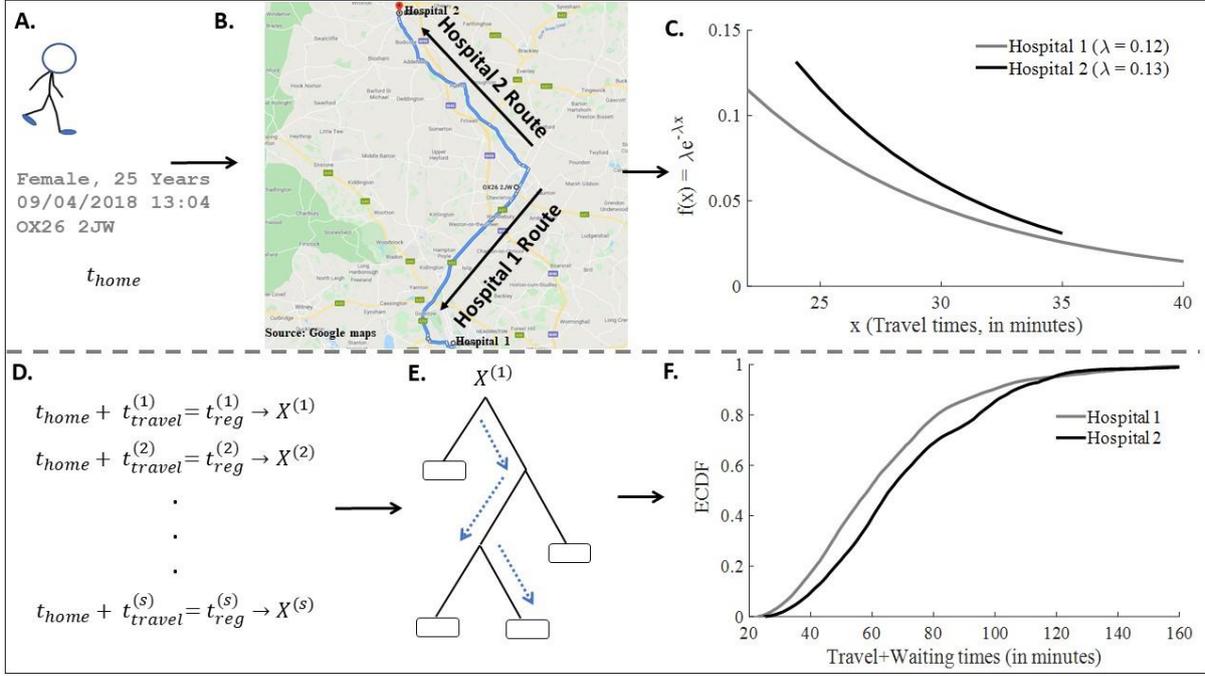

Figure 9 presents a schematic diagram that illustrates patient decision-making while selecting an ED site from a network with two hospitals. For example, consider a patient, say Anna, who needs to decide whether she should visit Hospital 1 or Hospital 2. For her location (postcode: OX26 2JW), and time of decision while at home (13:04 on 9 April 2018, say $t_{home}$), we access traffic information regarding the travel time to both hospitals. This information is used to estimate the probability distribution of travel times (Figure 9C). We draw a random sample of travel times (denoted by $t_{travel}^{(1)}, t_{travel}^{(2)}, ..., t_{travel}^{(s)}$) from the exponential distribution ($s$=500). For a given travel time, say $t_{travel}^{(1)}$, we estimate the corresponding time of registration $t_{reg}^{(1)}$ (as $t_{home} + t_{travel}^{(1)}$). Having estimated $t_{reg}^{(1)}$, we calculate the corresponding feature vector $X^{(1)}$ (Figure 9D), whereby $X^{(1)}$ is used as in input for QRF (Figure 9E depicts learning using one feature and one tree). For a given hospital site, feature vectors corresponding to different times of registration are used as inputs for QRF, to generate probabilistic forecasts of waiting times. Travel time estimates are added to the corresponding waiting time estimates, to compute the probability distribution of combined travel and waiting times (Figure 9F). In this example, using first order stochastic dominance to select between the distributions for the two hospitals, Anna would select Hospital 1. Her decision would contribute towards an increase in the overall patient-flow at Hospital 1. Since features are dynamically extracted for each patient, changes in load and congestion from Anna's decision would be reflected in the features for the next incoming patient at Hospital 1. Feedback of patient-decisions is thus included during the modelling.

It can be envisaged that providing information on estimates of travel and waiting can influence a patient's routing decisions, which can, in turn, affect the load (number of patients) and congestion (waiting times) at an ED. We study this impact by considering the following five alternative decision-making criteria for selecting between the two EDs: (1) shortest distance, (2) lowest mean travel time, (3) lowest mean of the probability distribution of the sum of travel and waiting times (*risk neutral* criterion), (4) lowest 75% quantile of the probability distribution of the sum of travel and waiting times (*risk averse* criterion), and (5) lowest 95% quantile of the probability distribution of the sum of travel and waiting times (*very risk averse* criterion). Table 5 compares the total load and congestion for the two hospitals (Hospital 1 and Hospital 2) for each of the five ED selection criteria. The simulation results suggest that selecting an ED based solely on the shortest distance criterion will result in diverting the vast majority of patients (291 out of 318) to Hospital 1, which would lead to exceedingly long waiting times compared to Hospital 2. Considering mean travel times only would also result in a highly nonuniform spread of load, whereby Hospital 2 would experience relatively high attendances and prolonged waiting times. Note that although the vast majority of the simulated postcodes were geographically closer to Hospital 1 on average (mean 14.0 miles) compared to Hospital 2 (mean 15.9 miles), the overall travel times to Hospital 1 were slightly longer (mean 28.9 minutes) than Hospital 2 (mean 26.2 minutes). Interestingly, selecting an ED based on either the lowest mean, 75% quantile, or 95% quantile of the combined travel and waiting times distribution results in a more uniform spread of patient load across the two hospitals, whereby the attendances and waiting times are similar for both hospitals, and no patient ends up waiting for exceeding long hours (>2 hours). Since Hospital 1 is a larger hospital with more resources compared to Hospital 2, the model typically ends up diverting more patients to Hospital 1. A larger load on the ED seems to be associated with longer waiting times in Table 5. We find that the empirical cumulative distribution function of waiting times are similar for the two EDs, in cases where patients consider both travel and waiting time estimates for decision-making. These results suggest that assisting patients in making informed routing decisions can potentially facilitate the uniform spread of load on EDs and help reduce waiting times, which is in broad agreement with previous findings (Dong et al. 2019).

In the absence of actual waiting times for both hospitals for a given patient (as a given patient can visit only one ED), point forecasts obtained from the QRF were treated as the true waiting times. Moreover, since it was not possible to estimate the time elapsed from arrival at the hospital to registration at the ED (e.g., time spent in the parking area), we treat the time of arrival at the ED as the actual time of registration for simulation. Thus, the numbers quoted in Table 5 should be treated with caution and be used only for comparative analysis across different criteria.

**Table 5.** Number of patients and their waiting times for the two hospital sites (Hospital 1 and Hospital 2) for the following five alternative ED selection criteria: shortest distance, lowest mean travel time, and lowest mean, 75%, and 95% quantile of the probability distribution of the sum of travel and waiting times.

| Load & Congestion | ED Selection Criteria | | | | |
|---|---|---|---|---|---|
| | Distance | Travel time | Mean *Risk Neutral* | 75% Quantile *Risk Averse* | 95% Quantile *Very Risk Averse* |
| $n_{Hospital\ 1}$ (low, medium, high) | 291 (45,168,78) | 71 (58,13,0) | 194 (158,36,0) | 188 (134,54,0) | 190 (154,36,0) |
| $n_{Hospital\ 2}$ (low, medium, high) | 27 (25,2,0) | 247 (6,108,133) | 124 (104,20,0) | 130 (114,16,0) | 128 (110,18,0) |
| $\mu_{t_{1,Hospital\ 1}}$ ($\sigma_{t_{1,Hospital\ 1}}$) | 93.1 (43.9) | 37.5 (15.6) | 36.1 (14.9) | 38.4 (16.3) | 36.2 (16.2) |
| $\mu_{t_{1,Hospital\ 2}}$ ($\sigma_{t_{1,Hospital\ 2}}$) | 22.8 (10.6) | 131.0 (49.8) | 32.6 (15.0) | 31.9 (13.5) | 32.1 (13.2) |

*Note*: $n_{Hospital\ 1}$ denotes total attendances for Hospital 1. The number of patients waiting for low ($<$ 45 minutes), medium (46 to 120 minutes), and high ($>$ 120) times are shown in brackets. The mean and standard deviation of waiting time for Hospital 1 are denoted by $\mu_{t_{1,Hospital\ 1}}$ and $\sigma_{t_{1,Hospital\ 1}}$, respectively. Similar notation is used for Hospital 2.

## 6. Summary and Concluding Remarks

In this study, we propose a machine learning approach using QRF to estimate the conditional quantiles of patient waiting times in an ED. The model utilized a rich set of features that were extracted from detailed patient-level records spanning five years. Rankings of predictor importance suggested that ED workload due to patient volumes and calendar effects were the most salient features. Model evaluation was based on an exhaustive comparison of distributional, quantile, and point forecast accuracy. Encouragingly, QRF convincingly outperformed the empirical benchmarks that are typically used in practice, along with the Q-Lasso and quantile regression methods that have been proposed in the literature for modelling waiting times. The performance of QRF was consistently superior for both hospital sites, which comprised 334,635 and 180,715 patient-level records. The consistency of our findings, using a large ED dataset for modelling waiting times, helps garner confidence in the generalizability of our results to data from independent ED sites.

Arrivals of low-acuity patients in the ED significantly increase the waiting times for high-acuity patients (Bayati 2017). This is an issue for high-acuity patients that need to be admitted, because delays in admitting a patient to the intensive care unit (ICU) has adverse effects on patient outcomes (Chan et al. 2017). In 2018, around 70% of US hospitals inpatients were processed via EDs (Augustine 2019). Thus, although we focus on modelling waiting times of low-acuity patients in EDs, this work could have implications for high-acuity patients and other parts of the hospital such as ICUs. The proposed modelling strategy could potentially be adapted to other applications, such as airports, railways, and call centres, where publishing waiting time estimates can help improve overall customer satisfaction.

A potentially useful line of future work would be to generate and evaluate patient waiting times conditional on the presenting complaint (e.g., cardiovascular disease, respiratory illness etc.). This would allow for more granular modelling of waiting times. The model estimates could be revised frequently to accommodate changes in the ED workload so that at any chosen moment, a patient could get an update on their waiting time distribution. Moreover, given that low-acuity patients are likely to drop out from EDs rather than wait in a crowded room for long hours, especially in times of social distancing, it can be envisaged that publishing waiting times of different service providers could be of particular benefit to patients and emergency healthcare services.